# QED in strong, finite-flux magnetic fields

M. P. Fry[*]

School of Mathematics, Trinity College, Dublin 2, Ireland

## Abstract

Lower bounds are placed on the fermionic determinants of Euclidean quantum electrodynamics in two and four dimensions in the presence of a smooth, finite-flux, static, unidirectional magnetic field $\mathbf{B}(\mathbf{r}) = (0, 0, B(\mathbf{r}))$, where $B(\mathbf{r}) \geq 0$ or $B(\mathbf{r}) \leq 0$, and $\mathbf{r}$ is a point in the $xy$-plane.

PACS number(s): 12.20.Ds, 11.15.Tk

[*]Electronic address: mpfry@maths.tcd.ie

Fermionic determinants, obtained by integrating out the fermionic degrees of freedom, produce an effective measure for the boson fields of a Euclidean field theory. They are at the heart of gauge field theories. In the case of quantum electrodynamics, which is our focus here, lack of nonperturbative information on the gauge invariant fermionic determinant has so far blocked a nonperturbative analysis of the theory in the continuum with dynamical fermions. Our purpose here is to supply lower bounds on the Euclidean determinants in two and four dimensions for the physically relevant case of smooth, finite-flux, static, unidirectional magnetic fields $\mathbf{B}(\mathbf{r}) = (0, 0, B(\mathbf{r}))$ subject to the constraint $B(\mathbf{r}) \geq 0$ or $B(\mathbf{r}) \leq 0$. Since $\nabla \cdot \mathbf{B} = 0$, $B(\mathbf{r})$ can only depend on the coordinates $x$ and $y$ which we denote by $\mathbf{r} = (x, y) \in \mathbb{R}^2$. After nearly half a century there are still no lower bounds on fermionic determinants in the literature that we are aware of for *any* inhomogeneous magnetic field except for the exactly solvable case of Euclidean $QED_2$ with a magnetic field confined to the walls of a cylinder passing through the $xy$-plane [1]. More will be said below about our choice of fields. Upper bounds on the determinants of $QED_2$[2,3] and $QED_4$[4] for these fields, without any constraint on the sign of $B$, are already known and will be restated below.

We will use Schwinger's proper time definition of the determinant of $QED_4$[5–7]:

$$\ln \det_{ren}(1 - S\!\!\!/\!A) = \frac{1}{2} \int_0^\infty \frac{dt}{t} \text{Tr} \left\{ e^{-P^2 t} - \exp\left[-(D^2 + \frac{1}{2}\sigma_{\mu\nu}F^{\mu\nu})t\right] + \frac{F^2}{96\pi^2} \right\} e^{-tm^2}, \quad (1)$$

where $D_\mu = P_\mu - A_\mu$; S denotes the free fermion Euclidean propagator; $m$ is the unrenormalized fermion mass, and the coupling, $e$, has been absorbed into the potential, $A_\mu$. This definition incorporates the second-order charge renormalization subtraction at zero-momentum transfer required for the integral to converge for small $t$; hence the determinant's subscript.

To proceed one has to specify $A_\mu$, which may be an external field or a random tempered distribution or a linear combination of both. The Gaussian measure, $d\mu(A)$, describing the continuum gauge field is concentrated on $\mathscr{S}'$, the space of tempered distributions. It is assumed that the gauge field is given a mass (an infrared cutoff) whose removal can be dealt with after integrating over the gauge field.

Since there is a need to regularize in both $QED_4$ and $QED_2$ one might consider replacing the random fields in the determinant and elsewhere in the functional integral, but not in the measure $d\mu(A)$ itself, with $A_\mu^\Lambda(x) = (h_\Lambda \star A_\mu)(x)$, where $A_\mu$ is convoluted with an ultraviolet cutoff function $h_\Lambda \in \mathscr{S}$, the functions of rapid decrease [2]. The potentials $A_\mu^\Lambda$ are polynomial bounded $C^\infty$ functions which produce the regularized photon propagator

$$\int d\mu(A)\, A_\mu^\Lambda(x)\, A_\nu^\Lambda(y) = D_{\mu\nu}^\Lambda(x-y), \tag{2}$$

where $D_{\mu\nu}^\Lambda$'s Fourier transform is such that $\hat{D}_{\mu\nu}^\Lambda \propto |\hat{h}_\Lambda|^2$, with $\hat{h}_\Lambda$ denoting the Fourier transform of $h_\Lambda$. One possibility is to choose $\hat{h}_\Lambda \in C_0^\infty$ with $\hat{h}_\Lambda(q) = 1$ for $q^2 \leq \Lambda^2$ and $\hat{h}_\Lambda(q) = 0$ for $q^2 > 2\Lambda^2$. In practice one may just assume that $A_\mu$ in Eq. (1) is polynomial bounded and $C^\infty$. If one succeeds in getting a useful gauge invariant result for $\det_{ren}$ one can always decide afterward precisely how to introduce an ultraviolet regulator before integrating over $A_\mu$.

At present we cannot deal with general fields $F_{\mu\nu}$ in $QED_4$. If we specialize to the case of static, unidirectional magnetic fields $(0, 0, B(\mathbf{r}))$, where $\mathbf{r}$ is a point in the $xy$-plane, we can make useful estimates provided $B(\mathbf{r})$ is further restricted to have finite flux.

The reason for this restriction is as follows. The relevant operator in Eq. (1) is now the supersymmetric Pauli Hamiltonian $H = (\mathbf{P} - \mathbf{A})^2 - \sigma_3 B \geq 0$. For polynomial, infinite-flux magnetic fields of the form $B = \sum_{n=0}^{N} \lambda_n (\mathbf{r} - \mathbf{c}_n)^{2k_n}$, where $\{\lambda_n\}$ and $\{\mathbf{c}_n\}$ are arbitrary real numbers and vectors, respectively, and $\{k_n\}$ are nonnegative integers, Avron and Seiler [8] have shown that the ground state of $H$ is infinitely degenerate and that the manifold of zero-energy bound-state wave functions is parameterized by a point in $\mathbb{R}^{2(2k_{max}+1)}$, irrespective of the translational invariance of the magnetic field. Thus, for the solved special case of $N = k_{max} = 0$ or $B = \text{const}$ [5,7,9], the ground state degeneracy, which persists in all excited states, induces a volume-like divergence in the determinant, and so one deals instead with the effective Lagrangian density. Neither the determinant as defined by Eq. (1) nor any other definition we are aware of can make sense out of such degeneracies. Hence our restriction to finite flux fields [10]. We will also assume $B(\mathbf{r})$ is square integrable.

Specializing to Euclidean $QED_2$ or equivalently, the massive Euclidean Schwinger model, we get from Eq. (1)

$$\frac{\partial}{\partial m^2} \ln \det_{Sch}(1 - S\!\!\!/A\!\!\!/) = \frac{1}{2} \text{Tr}\left[ (D^2 - \sigma_3 B + m^2)^{-1} - (P^2 + m^2)^{-1} \right], \quad (3)$$

where the charge renormalization counterterm in Eq. (1) is absent in $QED_2$. Next we introduce the following sum rule [11]:

$$\text{Tr}\left[ (D^2 - B + m^2)^{-1} - (D^2 + B + m^2)^{-1} \right] = \frac{\Phi}{2\pi m^2}, \quad (4)$$

where $\Phi = \int d^2 r\, B(\mathbf{r})$, and the trace is over space indices only. This sum rule follows from the invariance of $\det_{Sch}$ under the restricted gauge transformation $\varphi \to \varphi + \text{const}$, where in $1+1$ dimensions $A_\mu = \epsilon_{\mu\nu} \partial^\nu \varphi + \partial_\mu \chi$ and $B = -\partial^2 \varphi$. The left-hand side of Eq. (4) has been investigated by several authors [12] in connection with index theorems for the Dirac operator $\not{D}$ on non-compact Euclidean space-times. However, the full generality of Eq. (4) on $\mathbb{R}^2$ was not realized.

Applying the sum rule to Eq. (3) gives

$$\frac{\partial}{\partial m^2} \ln \det{}_{Sch} = \frac{\Phi}{4\pi m^2} + \text{Tr}\left[(D^2 + B + m^2)^{-1} - (P^2 + m^2)^{-1}\right], \tag{5}$$

where the trace is now over space indices only. Suppose $B(\mathbf{r}) \geq 0$. Because $A_\mu \in C^\infty$ and is polynomial bounded, the Combes-Schrader-Seiler (CSS) inequality [13] may be used to estimate the trace in Eq. (5):

$$\begin{aligned}
\text{Tr}\left[(D^2 + B + m^2)^{-1} - (P^2 + m^2)^{-1}\right] &\leq \frac{1}{(2\pi)^2} \int d^2r\, d^2p \left[(p^2 + B(\mathbf{r}) + m^2)^{-1} - (p^2 + m^2)^{-1}\right] \\
&= -\frac{1}{4\pi} \int d^2r \ln\left(1 + \frac{B(\mathbf{r})}{m^2}\right).
\end{aligned} \tag{6}$$

The CSS inequality relies on the Golden-Thompson-Symanzik [14] inequality in the form given by Ruskai [15], i.e. $\text{Tr}\left(e^{-(A+B)}\right) \leq \text{Tr}\left(e^{-\frac{B}{2}} e^{-A} e^{-\frac{B}{2}}\right)$, where $A$ and $B$ are self-adjoint and bounded below and $A + B$ is essentially self-adjoint on $\mathcal{D}(A) \cap \mathcal{D}(B)$, which happens to be the case here [13]. It also relies on Kato's inequality as extended by Simon [16]:

$$\left| e^{-(P-A)^2 t}(x, y) \right| \leq e^{-tP^2}(x, y), \tag{7}$$

for $x, y \in \mathbb{R}^n$.

Inserting Eq. (6) into Eq. (5) and integrating both sides from $m^2$ up to $m^2 = \infty$ gives

$$\ln \det{}_{Sch} \geq \frac{1}{4\pi} \int d^2r \left[B(\mathbf{r}) - \left(m^2 + B(\mathbf{r})\right) \ln\left(1 + \frac{B(\mathbf{r})}{m^2}\right)\right], \tag{8}$$

where we fixed the constant of integration by the condition that $\det{}_{Sch}(m^2 = \infty) = 1$. This is physically reasonable since the creation of virtual pairs becomes impossible when $m^2 = \infty$, thereby preventing the appearance of any nonlinear effects; it is also true order by order in a power-series expansion of $\det_{Sch}$. The case $B(\mathbf{r}) \leq 0$ is dealt with by replacing $B$ with $-B$ in Eq. (8).

Using $\ln(1+x) \leq x$ for $x \geq 0$, Eq. (8) gives the representative bound

$$-\frac{e^2 \|B\|^2}{4\pi m^2} \leq \ln \det_{Sch} \leq 0, \tag{9}$$

where we have explicitly introduced the coupling constant, and $\|B\|^2 = \int d^2r \, B(\mathbf{r})^2$. The upper bound in Eq. (9) is the "diamagnetic" bound [2,3], which is really an expression of the paramagnetic property of fermions. As mentioned above, the lower bound relies in part on Kato's inequality, which is an expression of the diamagnetic tendency of fermions when their spin is neglected. The competition between diamagnetism and paramagnetism has succeeded in placing a lower bound on $\det_{Sch}$ in the presence of $B$. It is perhaps surprising that all the nonlinearities in a power-series expansion of $\ln \det_{Sch}$ are bounded by a quadratic in the field strength.

We now proceed to use the bound (8) to place a lower bound on the fermionic determinant of $QED_4$. In connecting $QED_2$ with $QED_4$ the reader is reminded that $eB$ has the invariant dimension of $m^2$. It is intuitively clear that the dynamics of charged fermions in a unidirectional, static magnetic field must have something to do with *Euclidean* $QED_2$. The connection follows almost immediately after differentiating Eq. (1) with respect ot $m^2$ and is given by [4]

$$\begin{aligned}\ln \det_{ren} &= \frac{V_\|}{4\pi^2} \int \frac{d^2k}{(2\pi)^2} |\hat{B}(\mathbf{k})|^2 \int_0^1 dz \, z(1-z) \ln\left[\frac{k^2 z(1-z) + m^2}{m^2}\right] \\ &+ \frac{V_\|}{2\pi} \int_{m^2}^\infty dM^2 \, \ln \det_3(M^2),\end{aligned} \tag{10}$$

where $e$ has been absorbed into $B$ again; $V_\|$ is the volume of the $zt$ space-time box, and

$\det_3$ is defined by

$$\ln \det_{Sch} = -\frac{1}{2\pi} \int \frac{d^2k}{(2\pi)^2} |\hat{B}(\mathbf{k})|^2 \int_0^1 dz \frac{z(1-z)}{k^2 z(1-z) + m^2} + \ln \det_3. \tag{11}$$

The first term in (11) is the standard gauge invariant second-order vacuum polarization contribuition to $\det_{Sch}$. By definition, then, $\ln \det_3$ is the sum of all one-loop fermion graphs in two dimensions, beginning with the box graph since $C$ invariance is maintained by the proper time definition of the determinant. Combining Eqs. (11) and (8) we obtain

$$\begin{aligned}\ln \det_3(m^2) &\geq \frac{1}{2\pi} \int \frac{d^2k}{(2\pi)^2} |\hat{B}|^2 \int_0^1 dz \frac{z(1-z)}{k^2 z(1-z) + m^2} \\ &+ \frac{1}{4\pi} \int d^2r \left[ B(\mathbf{r}) - (m^2 + B(\mathbf{r})) \ln\left(1 + \frac{B(\mathbf{r})}{m^2}\right) \right].\end{aligned} \tag{12}$$

Suppose $B$ is sufficiently strong that $\|B\|^2 \geq m^2$. Then we can split the mass integral in Eq. (10) into an integral from $m^2$ to $\|B\|^2$ and from $\|B\|^2$ to $\infty$ and insert (12) into the first integral. This gives

$$\begin{aligned}\ln \det_{ren} &\geq \frac{V_\|}{4\pi^2} \int \frac{d^2k}{(2\pi)^2} |\hat{B}|^2 \int_0^1 dz\, z(1-z) \ln\left[\frac{k^2 z(1-z) + \|B\|^2}{m^2}\right] \\ &+ \frac{V_\|}{8\pi^2} \left\{ \frac{\Phi}{2} (\|B\|^2 - m^2) - \int d^2r \left[ \|B\|^2 \left(\frac{1}{2}\|B\|^2 + B(\mathbf{r})\right) \ln\left(1 + \frac{B(\mathbf{r})}{\|B\|^2}\right) \right.\right.\\ &\left.\left. + \frac{1}{2} B^2(\mathbf{r}) \ln\left(\frac{\|B\|^2 + B(\mathbf{r})}{B(\mathbf{r}) + m^2}\right) - m^2 \left(\frac{m^2}{2} + B(\mathbf{r})\right) \ln\left(1 + \frac{B(\mathbf{r})}{m^2}\right) \right] \right\} \\ &+ \frac{V_\|}{2\pi} \int_{\|B\|^2}^\infty dM^2 \ln \det_3(M^2).\end{aligned} \tag{13}$$

We can estimate the last term in (13) in the limit of a strong magnetic field. Letting $A_\mu \to \lambda A_\mu$, with $\lambda > 0$, we obtain the bound

$$\ln \det_{ren} \underset{\lambda \gg 1}{\geq} \frac{V_\| \|B\|^2}{48\pi^2} \lambda^2 \ln \lambda + O(\lambda^2), \tag{14}$$

where [4]

$$\int_{\lambda^2 \|B\|^2}^\infty dM^2 \ln \det_3(M^2, \lambda B) \underset{\lambda \gg 1}{\sim} O(\lambda^o). \tag{15}$$

Before turning to Eq. (14) a comment on Eq. (15) is in order. As explained in [4], we can go back to the definition of $\ln\det_{Sch}$, Eq. (1) with the charge renormalization subtraction omitted, and make a heat kernel expansion since (15) requires the large mass limit and $m^2 \to \infty$ implies small $t$. Not surprisingly, the leading term in $\ln\det_3$ is related to the Euler-Heisenberg effective Lagrangian. We found that

$$\ln\det_3 \underset{m^2\to\infty}{\sim} \frac{e^4 \int d^2r \, B^4}{90\pi m^6} + O\left(\frac{e^4 \int B^3 \nabla^2 B}{m^8}\right), \tag{16}$$

from which (15) follows. As a check on (16) it gives, according to Eq. (10), a contribution to $\ln\det_{ren}$ from the box graph of $2\alpha^2 V_\parallel \int B^4/45m^4 + O\left(\alpha^2 \int B^3 \nabla^2 B/m^6\right)$. This gives the leading term of the effective Lagrangian for a constant magnetic field, i.e. $\mathcal{L}_{eff} = 2\alpha^2 B^4/45m^4 + O(\alpha^3)$, in agreement with previous results [5,7,9].

Finally, we can combine the lower bound (14) with the upper bound in [4] and state our final result in the form

$$\frac{e^2 \|B\|^2 V_\parallel}{48\pi^2} \leq \lim_{\lambda\to\infty}\left(\frac{\ln\det_{ren}(\lambda B)}{\lambda^2 \ln\lambda}\right) \leq \frac{e^2 \|B\|^2 V_\parallel}{12\pi^2}. \tag{17}$$

Unlike $QED_2$, the "diamagnetic" inequality (the upper bound in Eq. (9)) fails in $QED_4$ [6] due to the charge renormalization counterterm present in Eq. (1). Its offending sign is directly related to the fact that $QED_4$ is not asymptotically free. This fact and the opposing paramagnetic property of charged fermions combine in a way that cause $\det_{ren}$ to grow faster than an inverted Gaussian for the class of strong magnetic fields we have considered.

Bounds on the fermionic determinant of Euclidean $QED_3$ may also be obtained by means similar to those above and will be the subject of a future paper.

References and Footnotes